\documentclass[12pt]{spieman}  
\usepackage{amsmath,amsfonts,amssymb}
\usepackage{graphicx}
\usepackage{setspace}
\usepackage{tocloft}
\usepackage{ulem}
\usepackage[us,12hr]{datetime} 

\def\etal{et\ al.}
\def\deg{$^o$}
\def\mnras{MNRAS}
\def\aap{A\&A}
\title{Sub-kilometre scale ionospheric studies at the SKA-Low site, using MWA extended baselines}

\author[a,b,c]{Mar\'{\i}a  J. Rioja}
\author[b]{Richard Dodson}
\affil[a]{CSIRO Astronomy and Space Science, PO Box 1130, Bentley WA 6102, Australia}
\affil[b]{ICRAR, M468, The University of Western Australia, 35 Stirling Hwy, Crawley, Western Australia, 6009}
\affil[c]{Observatorio Astron\'omico Nacional (IGN), Alfonso XII, 3 y 5, 28014 Madrid, Spain}

\cftpagenumbersoff{figure}
\cftpagenumbersoff{table} 
\begin{document} 
\maketitle



\begin{abstract} 
The ambitious scientific goals of SKA require a matching capability for calibration of instrumental and atmospheric propagation contributions as functions of time, frequency and position. 
The development of novel calibration algorithms to meet these requirements is an active field of research.
In this work {we aim to characterize} these, focusing on the spatial and temporal structure scales of the ionospheric effects; ultimately, these provide the guidelines for designing the optimum calibration strategy. 
We used empirical ionospheric measurements at the site where the SKA-Low will be built, using MWA Phase-2 Extended baseline observations and the station-based Low-frequency Excision of Atmosphere in Parallel ( LEAP) calibration algorithm.
We have done this via direct regression analysis of the ionospheric screens and by forming the full and detrended structure functions.
We found that 50\% of the screens show significant non-linear structures at scales $>$0.6km that dominate at $>$2km, and 1\% show significant sub-minute temporal changes, providing that there is sufficient sensitivity.
Even at the moderate sensitivity and baseline lengths of MWA, non-linear corrections are required at 88 MHz during moderate-weather and at 154 MHz during poor weather, or for high SNR measurements.
Therefore we predict that improvements will come from correcting for higher order defocusing effects in observations with MWA Phase-2, and further with new developments in MWA Phase-3. 
Because of the giant leap in sensitivity, the correction for complex ionospheric structures will be mandatory on SKA-Low, for both imaging and tied-array beam formation.

\end{abstract}


{\noindent \footnotesize\textbf{*}Mar\'{\i}a  J. Rioja,  \linkable{maria.rioja@icrar.org} }



Compiled on \today\ at \currenttime

\section{Introduction}
\label{sect:intro}  

Low frequency radio astronomy is under-going a rebirth, with the arrival of the next-generation instruments that will provide two orders of magnitude improvement in sensitivity and an order of magnitude improvement in frequency coverage. In particular the Low frequency Square Kilometre Array (SKA-Low) will provide an instantaneous frequency coverage of 50 to 350 MHz, a collecting area of 0.4 sq. kilometres, spread over 65km baselines in the Phase-1 and the full 1.0 sq. kilometres in Phase-2, and exciting prospects of new science.
%
Thus it is vital that we update the radio-astronomical methods and strategies to match the potential of the radio-astronomical instruments. 
In this paper we focus on the challenges at the foundation of accurate direction-dependent ionospheric calibration, as these are 
among the most difficult obstacles to achieving the nominal SKA-Low performance. 

Directional dependent (DD) calibration is vital at low frequencies, because the field of view (FoV) is intrinsically large and the 
propagation effects caused by inhomogeneties in the distribution of the ionospheric plasma ($\Delta{\rm I}$, measured in TECU or 10$^{16}$ electrons per m$^2$) imposes temporal and spatial phase disturbances on the incoming wavefront whose magnitude scales inversely proportional to the observing frequency ($\phi_{\rm ion}({\rm deg})=480\,\Delta{\rm I}\, \nu_{\rm GHz}^{-1}$,  with $\nu_{\rm GHz}$ the observing frequency in GHz \cite{sovers_98}). 

%
Traditionally, with observations of smaller fields of view and at higher frequencies, these effects could largely be handled with a single station-based correction (or direction independent (DI) correction), valid for all of the field of view. 
Lonsdale (2005, Figure 1)\cite{lonsdale_05} classified four operating regimes based on the relative sizes of the FoV (V), the ionospheric phase fluctuation disturbances (S) and the projected size of the array (A). 
Traditionally, with observations of smaller fields of view and at higher frequencies (V$\ll$S), these effects could largely be handled with a single correction per station  valid within the FoV. This applies for both compact (A$\ll$S) and extended (A$\gg$S) arrays, which are cases 1 and 2, respectively, for which conventional self-calibration techniques are suitable.

For modern low-frequency arrays this is no longer sufficient, as the FoV is greater than the scale size of the disturbances (V$\ge$S).
Case 3 is for sufficiently compact arrays such that the DD wavefront disturbances over the array can be treated as planar (i.e. linear approx.) and yield coherent apparent source position shifts varying in time and direction. When the linear approximation is no longer valid (i.e. extended array) the presence of non-planar (i.e. curvature) disturbances additionally lead to source shape deformations; this is case 4, which requires a more complex calibration algorithm, where an independent solution for each direction and station must be applied.
We show here that this is an important issue even for the pathfinders such as Murchison Widefield Phase-2 Array (MWA-2), which has baselines less than 5 kilometres long.

A prime challenge with DD corrections is in the imaging of the data. The Van-Citter relationship that is the basis of all radio-interferometric imaging requires that the data in the Fourier domain has a uniform calibration applied before transforming it to the image domain. 
Several research groups are developing solutions for this issue, for example DDFacet\cite{ddf_tasse}, which takes a facetting approach, and is widely used for LOFAR LBA imaging and WSClean\cite{offringa_14} with Image Domain Gridding\cite{tol_18}, which uses gridding in the image domain to correct for direction-dependent effects, and is widely used for MWA imaging.

We have been looking at efficient methods to generate accurate DD-calibrations, for application in these next generation imaging tools. 
We have been developing a method called LEAP (Low-frequency Excision of Atmosphere in Parallel), which is discriminated from other methods by the fact that every direction can be treated as independent, and thus analysed in parallel. This compares favourably to sequential (`peeling') methods (e.g. SPAM\cite{intema_09} or RTS\cite{mitch_cal}) that correct for each source in order of strength, and simultaneous methods (e.g. SAGECal\cite{sagecal}) that solve for all directions with a very large solution matrix, which do not scale well to large numbers of sources or stations. 

LEAP of course has its limitations; the fact that it
does not perform source subtraction can elevate the noise floor, and the flux density cutoff for calibrator sources. 
Nevertheless, as it does not require a complete sky model (other than a catalogue of LEAP calibrator positions, which are each treated independently) and is very 
simple and robust, it can play an important role in providing a preliminary solution to the more sophisticated and complicated methods, and a unique role 
in providing the real-time direction-dependent calibration for the SKA tied-array beamforming. The latter is used to form multiple simultaneous beams within 
the field of view, as required in VLBI\cite{skavlbi_wp10} and equally in pulsar studies. 

Traditional MWA DD-calibration strategies, as used by GLEAM\cite{nhw_gleam} for example, 
assume that observations fall in ``Case 3'', and ignore the potential defocusing effect arising from non-linear disturbances in the wavefront above the array. The ionospheric calibration is carried out in the image domain, as a rubber sheet-type correction to the individual 2-min long scan images before mosaicing. 
This approach worked perfectly well for GLEAM (with a maximum MWA Phase-1 baseline of 2\,km), and has had some success with GLEAM-X (with a maximum MWA Phase-2 baseline of 5\,km). 
However, as we will show, for the increased sensitivity 
of SKA-Low this is going to be completely inadequate, even for this range of short baseline lengths.  


The ionospheric calibration data itself provides a means to gain information on the physical nature, and spatial and temporal structure scales of the propagation media
\cite{mevius_16}.
Previous MWA large-scale studies have been performed using image-based analysis \cite{loi_15,jordon_17,helmboldt_20}, however these cannot probe the fine scale.

In this paper we use the LEAP station-based calibration data, in the visibility domain, to quantify the presence of higher order spatial and temporal phase structure over the MWA Phase-2 array imprinted in the incoming wavefront as it propagates through the ionospheric plasma irregularities, for directions within the FoV.
As LEAP measures the phase variations within the baselines formed by the array, we are able to study the small scale spatial and temporal structures; that is at scales from 30m to 5km.  Our particular interest is to discover at what level the planar distortions approximations breakdown, entering the calibration regime 4, as that is crucial information for the planning of the SKA-Low calibration schemes. 

\subsection{A LEAP primer}

We quickly recap the discussions in Rioja et.\,al., 2018 \cite{rioja_18} (hereafter LEAP-I) 
to summarise the operations of LEAP.
The core insight of the LEAP algorithm is that the extremely wide fractional frequency bandwidth of current and further low-frequency arrays allows the use of frequency smearing to suppress all sources away from the phase centre.
LEAP processing only requires as input the directions for calibrator sources that are sufficiently strong to dominate the phase observable 
after the other directions are suppressed by frequency averaging. For this we have used the GLEAM-I catalogue\cite{nhw_gleam}, so the positions are accurate for the frequencies being used. This simple approach is remarkably effective, as the contributions from other directions are suppressed by a factor $\propto {{\Delta\nu}\over{\nu}} \theta_0 $, where $\Delta\nu$ is the bandwidth and $\nu$ is the central frequency, and $\theta_0$ is the source distance from the centre in beamwidth units. 
The fractional bandwidth is very high for modern low-frequency instruments. SKA-Low, for example, has a frequency span ratio close to 1; MWA has one as high as 1:3 and MWA Phase-3 (MWA-3) will be even higher. 
We require the input data to be, at a minimum, bandpass corrected so that the frequency channels can be averaged. For example, 
if using {\it a priori} conventional direction independent (DI) self-calibration (i.e. a flux-weighted average solution across the FoV), LEAP  will correct 
for the DD ionospheric residuals with respect to the flux-weighted DI calibration direction; these DD residuals being small we can then determine the DD dispersive delay unambiguously from the phase term only. 

\begin{figure} 
    \centering
    \includegraphics[width=0.75\textwidth]{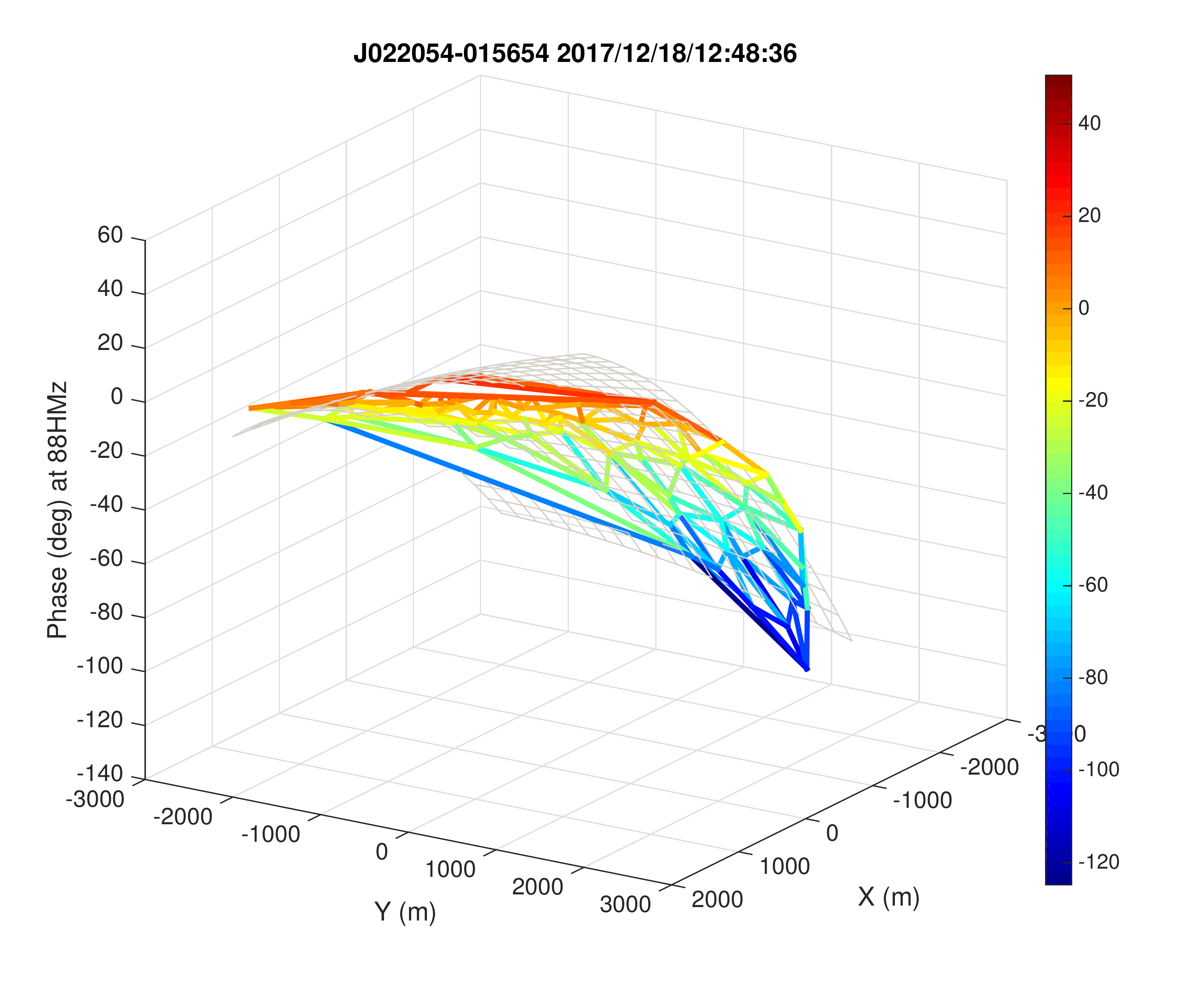}
    \caption{
    An example of a LEAP ionospheric phase screen at 88MHz from a 2-min scan considered in this paper; the span in the sky is 30$^\prime$-60$^\prime$, depending on the height of the ionosphere.
    Shown as a mesh are the station-based LEAP calibration phases above the MWA Phase-2 extended configuration array for the direction of one LEAP calibrator along a line of sight with an elevation of 65\deg, color-coded in degrees, with the X and Y axis in meters for the 128 stations (at the nodes of the mesh).     
    The wireframe shows the second order fit to the data, underlining the high curvature of the surface. 
    \label{fig:example_phase}}

\end{figure}

An example of the phase screens ({\it hereafter ionospheric screen}) over the array observed with the MWA-2, for a given direction, is presented in Figure \ref{fig:example_phase}. 

\begin{figure}
    \centering
    \includegraphics[width=0.85\textwidth]{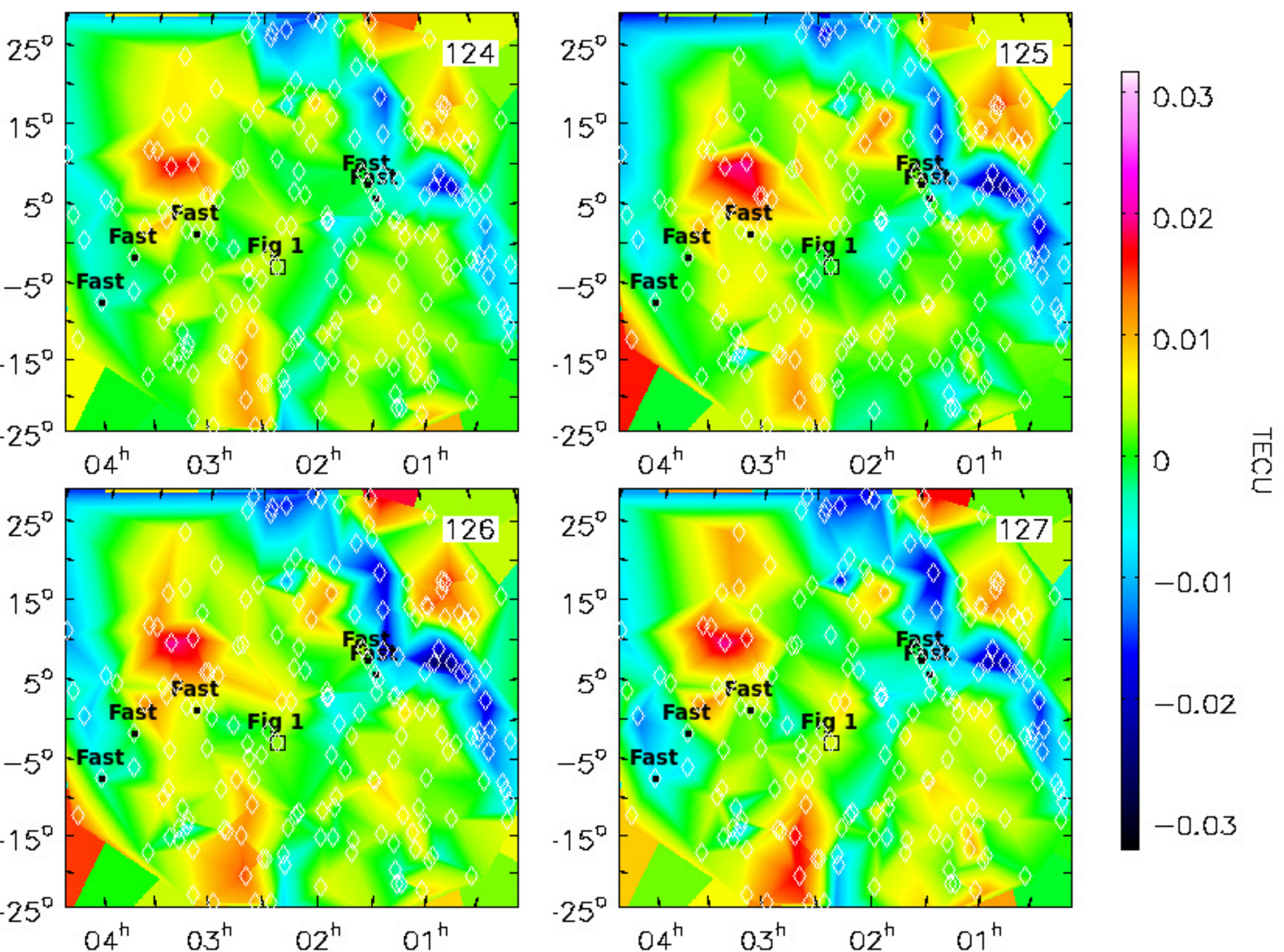}
    \caption{
    Example of the ionospheric phase screens as would be provided to WSClean for DD wide-FoV imaging, from a 2min scan at 88\,MHz in our dataset. Each plot represents the ionospheric phase solutions for a given station across the wide MWA FoV, for four stations in the outer region of the extended configuration (i.e. long baselines).
    The TEC values have been interpolated from the LEAP station-based phase solutions towards the $\sim$200 calibrator directions, which are shown with white diamonds. The location of calibrator shown in Figure \ref{fig:example_phase} is marked with a black square (labelled Fig. 1) and the location of the few fast changing sources (labelled Fast) are marked with a star. 
    \label{fig:WSC_corrections}}
\end{figure}

The interpolation and smoothing of these solutions over the array, and over different pointings, provides an input into, for example, WSClean for direction dependent calibration and imaging. An example is shown in Figure \ref{fig:WSC_corrections}, where the interpolated TEC surface for all directions on the sky is projected onto a regular grid, for each station. 

\section{Observations and Analysis}\label{sec:obs_ana}

We used two datasets of GLEAM\cite{nhw_gleam,gleam-x} observations with the MWA Phase-2 long baseline configuration\cite{mwa2} at two of the lowest frequency bands, with an instantaneous bandwidth of 30 MHz for the explorations presented for this paper.

One dataset consisted of a series of interleaving 2-min long scans, centred at 88MHz and 154MHz, on December 
18th, 2017, under moderate-weather conditions.
The interval between observations at the same frequency was 10 minutes, spanning an hour and a half.

The other dataset includes a small number of 2-min long scans carried out on June 2018, with the strong source 
3C444 (J221425-170140) near the phase centre, under a variety of weather conditions. 
That is, good weather on June 3$^{\rm rd}$ and poor weather on June 12$^{\rm th}$\cite{nhw_pc}.
The latter was flagged to be rejected and reobserved based on severe coherent artefacts (residual sidelobes from several strong sources caused by ionospheric distortions) 
that contaminated the whole of the image, following the standard GLEAM analysis.

All the observations were calibrated using LEAP, following the calibration and imaging procedure described in Section 1.1, as part of a wider-scope project to advance the effective imaging implementation (e.g. a single gridding and imaging inversion) of LEAP DD calibration
along with a comparative study of `refocusing' calibration in wide FoV imaging using MWA 
observations (Dodson \etal, 2021, in prep).

However the aim of this paper is to provide answers to some of the questions about the local spatial and temporal scales of the ionospheric disturbances pertinent for the SKA, and identify the requirements for an optimum strategy for the removal of systematic errors 
depending on the  observing frequency and under different ionospheric conditions.




Of special interest for this work are the LEAP calibration by-products, namely the measured station-based phase solutions which define the so-called ionospheric phase screens above the array, along the directions of the stronger LEAP calibrators 
(i.e. those with reduced thermal noise error contribution). They are a measure of the wavefront disturbances introduced by its propagation through the ionosphere.
Hence, each screen conveys information of the spatial (and temporal) structure of the DD ionospheric disturbances, at scales smaller than the array size 
(i.e. less than 5km and potentially down to 30m). They are the main observable for the ionospheric analysis in the visibility domain followed here.

The station-based phase observables are obtained with the script discussed in LEAP-I, where {\sc casa} was used to: call WSClean for phase rotation to calibrator source directions and imaging for analysis and validation, and to solve for gain solutions. 
As part of ICRAR bridging contributions to the SKA science processing a GPU version has been written\footnote{https://jira.skatelescope.org/browse/SP-994} that can efficiently process multiple directions in parallel. This addresses the issue of parallel reading of the data, however the processing here predated the release of that tool.

We have followed two methods of analysis of the individual ionospheric phase screens. 
Firstly, a regression analysis fitting, using linear and second order surfaces.


As a result we measured the planar `slope' and orientation from the 1$^{\rm st}$ order 2D linear fits, along with the `curvature'; obtained as 
the quadratic mean of the three 2$^{\rm nd}$ order fits coefficients.

Also we measured the Root Mean Squared residual errors (RMSE) to the first and second-order fit (${\rm RMSE}_1$ and ${\rm RMSE}_2$ respectively), and computed the RMSE fractional improvement, defined, for a given screen, as:
\[
\Delta {\rm RMSE}=\frac{{\rm RMSE}_1 - {\rm RMSE}_2}{ {\rm RMSE}_1}
\]

Secondly, we compute the 2nd-moment structure function ({\it hereafter} Full Structure Function or SF) of the ionospheric screens ($\phi_{ion}$) over the range of (projected) baselines in the MWA array, $r$, following the approach in, for example, Van de Tol, 2009\cite{vantol_phd}:
\[
D(r)=\langle (\phi(r^\prime)-\phi(r^\prime+r))^2\rangle
\]
were $\langle ....\rangle$ stands for a statistical average.


The SF analysis is traditionally used to measure the scaling features of the electron density fluctuations in the ionosphere in order to understand the nature of the physical processes (e.g. turbulences) at the origin of the measured signal. 

Here we use the ability of the SF as a tool to characterize complex surfaces, particularly with non-uniformly sampled and non-periodic data, using the measured phase values that form the ionospheric screens.
\cite{mevius_16,nature_sf_swarm,sf_toys_neuroscience}. Of particular interest here is the capacity to detect deviations from linearity.
We calculated the SFs using the baseline phase values, deduced from the station solutions. 
In addition we formed the Detrended Structure Function (DSF) from the residual phase values, after subtracting the corresponding planar fits from the station solutions. 
This eliminates the overall dominant influence of a planar slope in the data set.
%

We expected D(r), in a noise-free environment, to follow: 
\[
D(r) = C^2 r^\beta
\]
where $C$ would be the slope of the ionospheric phase screen (i.e. change of phase per unit baseline distance), and 
where $\beta$ is the so-called scaling exponent, which allows us to characterize the scaling nature of the signal under investigation (in our case, the ionospheric disturbances); this value is equal to 5/3 for pure thin-screen Kolmogorov turbulence, or 2 for perfectly planar surfaces. This behaviour holds over the inertial region, which is the approximately linear regime in log-log space between the outer scale at which power is injected and the inner scale at which power is dissipated in the Kolmogorov theory of turbulence.

We also generated spatial structure functions using simulated datasets of synthetic phase surfaces base on our MWA observations, i.e ionospheric toy-models with known properties, to test the applicability of SF analysis to the characterization of ionospheric phase screens
and to help with interpretation of our empirical findings in terms of ionospheric behaviour. 
The toy models consisted of a family of surfaces defined by 2D linear and quadratic polynomial functions with and without added noise, 
for a range of values for the polynomial coefficients and noise parameters, compatible with our findings from the regression analysis.

Dodson \etal{} (2021, in prep.) will present a complementary analysis of the ionospheric disturbances based on the comparison of pre- and post-calibration images. 
The latter has the capacity to additionally correct for defocusing artifacts in the image, such as changes in the source shape and residual sidelobe patterns, which arise from the small scale ionospheric disturbances.

\section{Results}


\begin{figure} 
    \centering
    \includegraphics[width=0.75\textwidth]{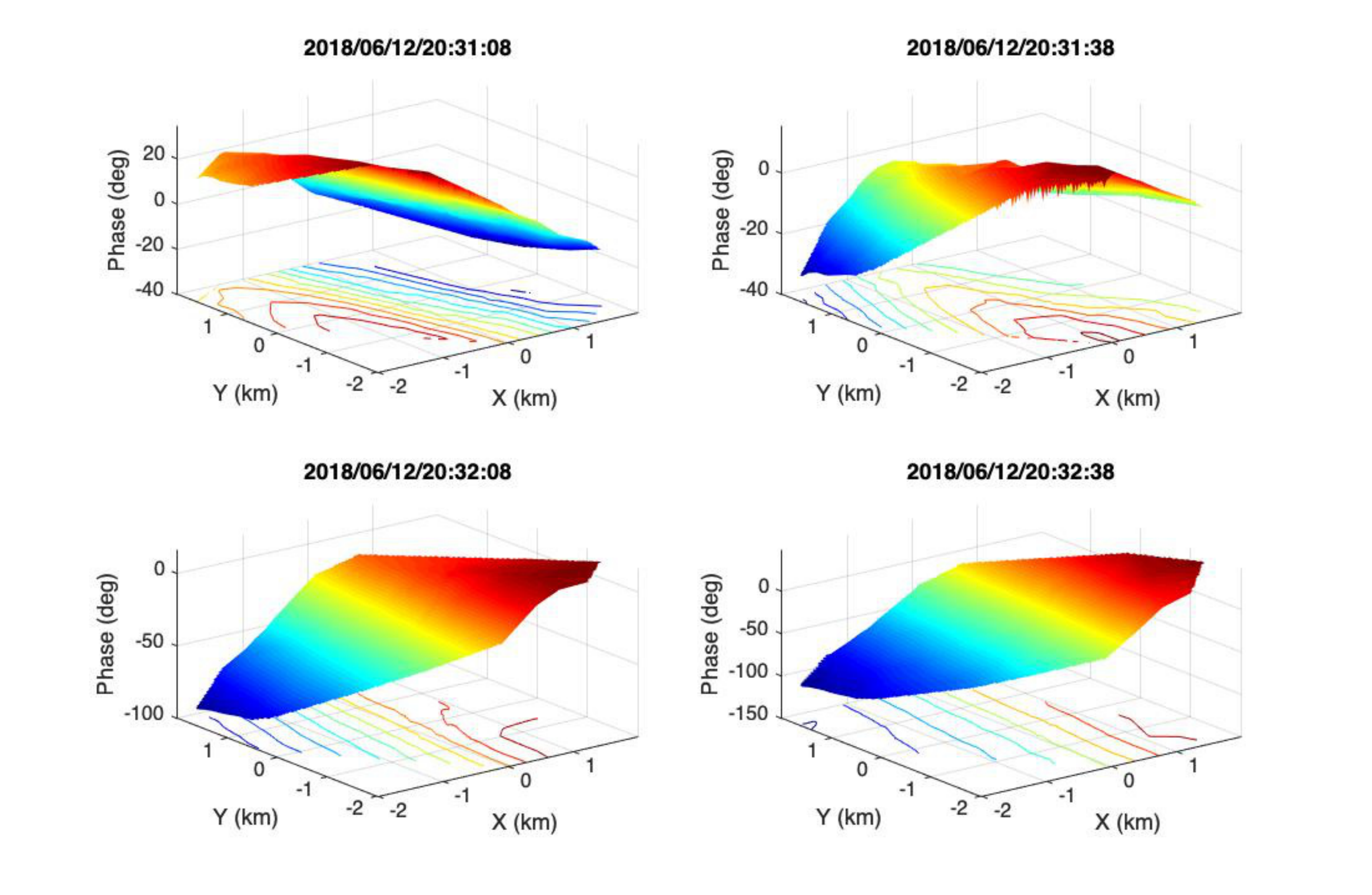}    
    \caption{ 
    A highly time variable phase screen, with 30s sampling, observed on a nominally good weather day (2018/06/12) on 3C444 at 154MHz. A compact ($<$1km) knot of plasma can be seen passing through the line of sight of the array. 
    About 1\% of the sources with SNR$>$6 have high temporal variability. 
    \label{fig:example_phase_temporal}}
\end{figure}

Figure \ref{fig:example_phase} showed an example of LEAP ionospheric phase screens above the MWA array in the direction of a strong LEAP calibrator, for a 2-min scan solution. 
For weaker sources, it is expected that the ionospheric signature will be diluted, and ultimately buried under the increasing thermal noise. The results presented here comprise the analysis of the ionospheric screens in the directions of all ($\sim$200) LEAP calibrators (with SNR$>$2, per station) within the MWA FoV in each 2-min scan, for the set of scans and the analysis described in Section \ref{sec:obs_ana}.

{\it Regression Analysis}

\begin{figure}
    \centering
    \includegraphics[width=0.7\textwidth]{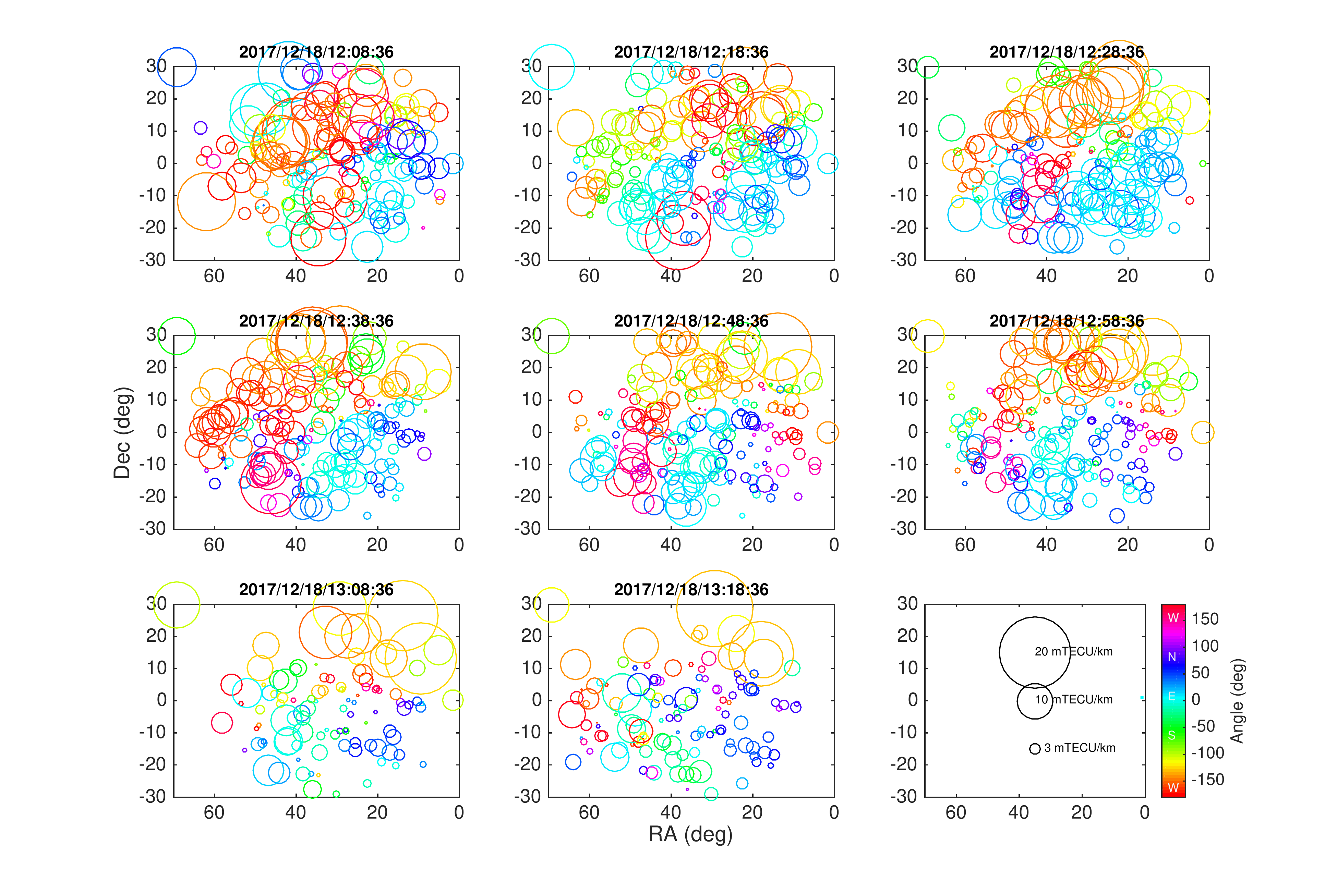}
    \includegraphics[width=0.7\textwidth]{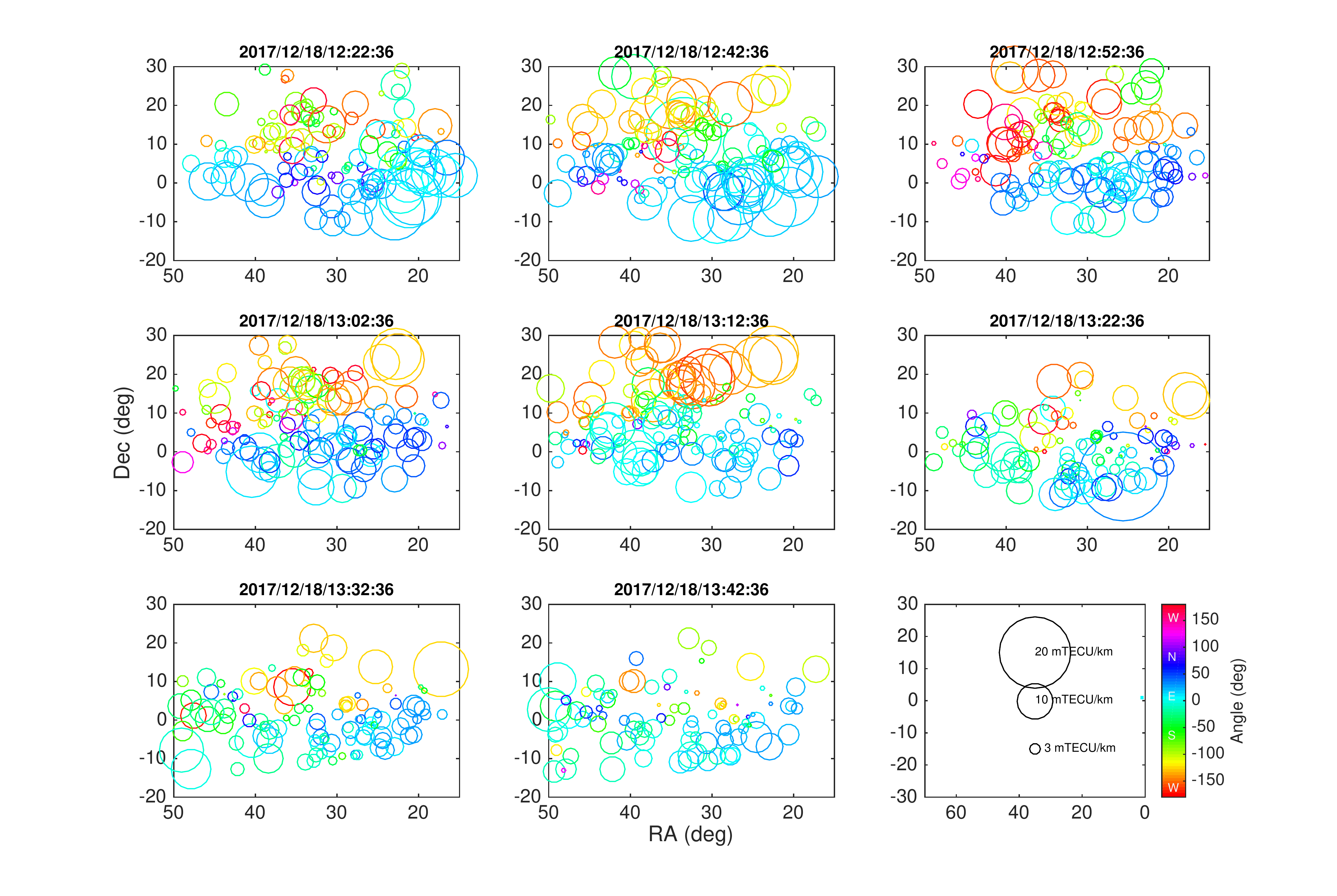}
    \caption{
    Outcomes of 2D linear regression analysis of LEAP ionospheric screens: fitted values of slope (circle size, see last frame) and orientation angle (East through North, cyclic colour scale}) across the array, 
    for the RA and DEC directions of LEAP calibrators with SNR$>$2 per station. {\it Upper} panel for 88 MHz, and {lower} panel for 154 MHz. Each subplot is for a 2-min scan at the time indicated in the title. The results for any snapshot at either frequency are spatially correlated, even though the ionospheric screens in each direction are measured independently.
    \label{fig:Corrections}
\end{figure}

Figures \ref{fig:Corrections}a) and b) show the outcomes from the linear 2D regression analysis of the unwrapped station-based phase solutions in an ionospheric screen, for 8 consecutive 2-min scans, at 88 and 154 MHz, respectively.
Unwrapping, that is taking into account full wraps of phase by enforcing continuity, is essential to accurately fit the phase slopes on the longer baselines.
Shown are the magnitude of the `slope' (circle size, mTECU/km) and the orientation angle (circle colour, degrees) of the fitted plane.
The sky location of the symbols correspond to that of the LEAP calibrators across the FoV, for each scan. 
The FoV at 154\,MHz is smaller than for 88\,MHz, and each subplot is taken at a different moment in time (although all those shown have a common phase centre). Thus the station gains vary between each subplot and the selected sources vary between scans.

\begin{figure}
    \centering
    \includegraphics[width=0.7\textwidth]{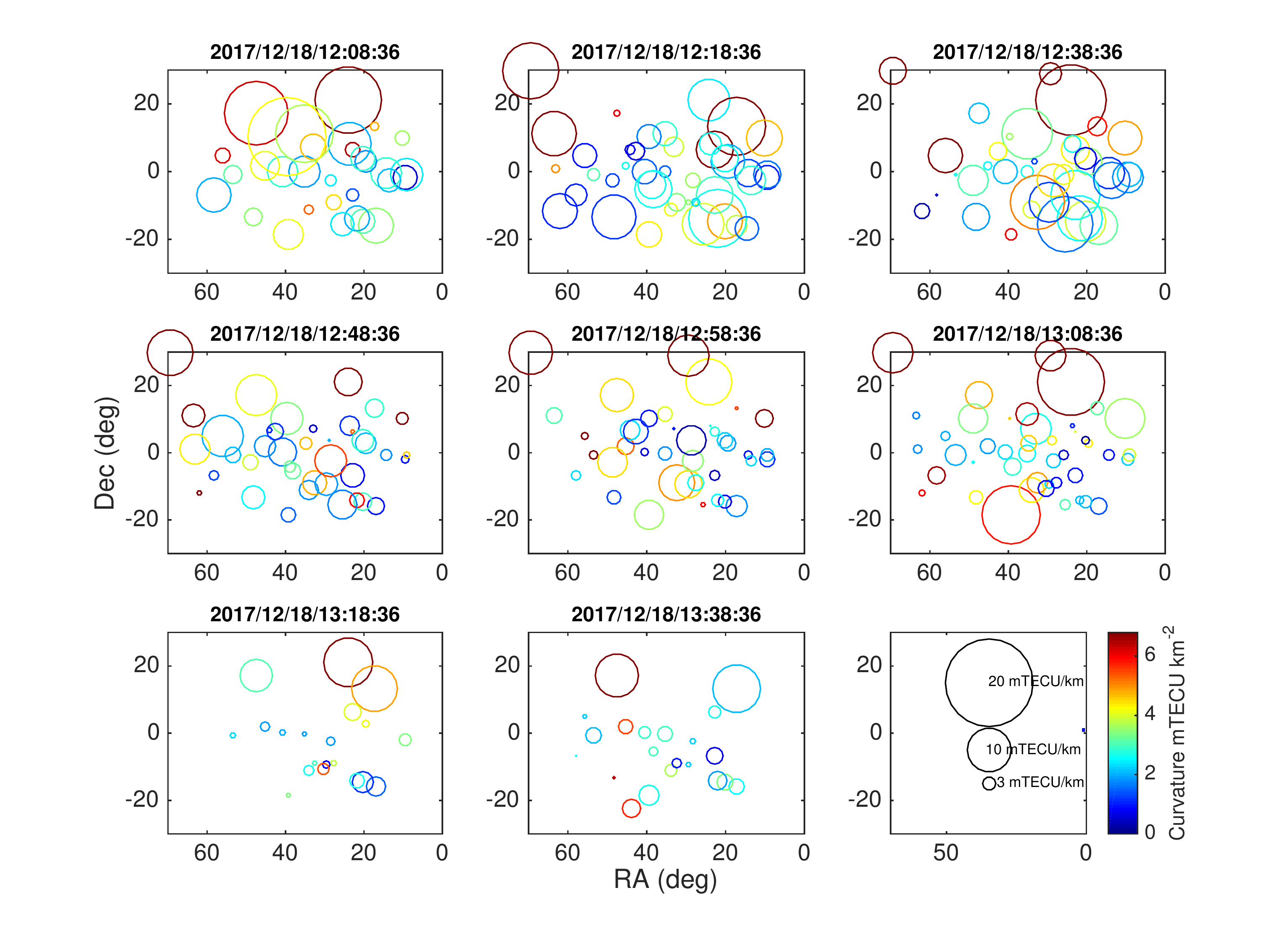}
    \caption{
    Outcomes of 2D second-order regression analysis of LEAP ionospheric screens: fitted values of slope (circle size, see last frame and identical to Figure \ref{fig:Corrections}) and curvature (colour) across the array, 
    for the RA and DEC directions of LEAP calibrators with SNR$>$6 per station, at 88 MHz. Each subplot is for a 2-min scan at the time indicated in the title. The curvature shows significantly less spatial correlation than the slope.}
    \label{fig:Curved_corrections}
\end{figure}

Figure \ref{fig:Curved_corrections} corresponds to the second order 2D regression analysis, 
showing the magnitude of the fitted curvature (circle colour, mTECU/km$^{2}$) and slope (circle size with same scale as Figure \ref{fig:Corrections}, in mTECU/km), at 88\,MHz, with SNR$>$6. 
Because higher SNR is required to measure the curvature, there are significantly fewer data points. Furthermore, we do not show the corresponding figure for 154\,MHz; there were too few data points per scan to show the spatial correlation.

The median value for the ionospheric slope is 5$\pm$3\,mTECU/km, across the $\sim$5km span of the array, for all LEAP calibrator directions, at both 88 and 154 MHz. 
The median value for the curvature is 3$\pm$2\,mTECU/km$^2$, at both frequencies for SNR$>$6.

For each screen, the linear and second order RMSE errors from the regression analysis are combined to form the fractional $\Delta$RMSE quantity, that would show
a significant improvement of one fit above the other, as expected if deviations from linearity are significant.
That is, cases when the curvature signature is large enough so that its magnitude is comparable to or greater than the thermal noise contribution and results in a large fractional reduction of the RMSE.

\begin{figure}
    \centering
    \includegraphics[width=0.75\textwidth]{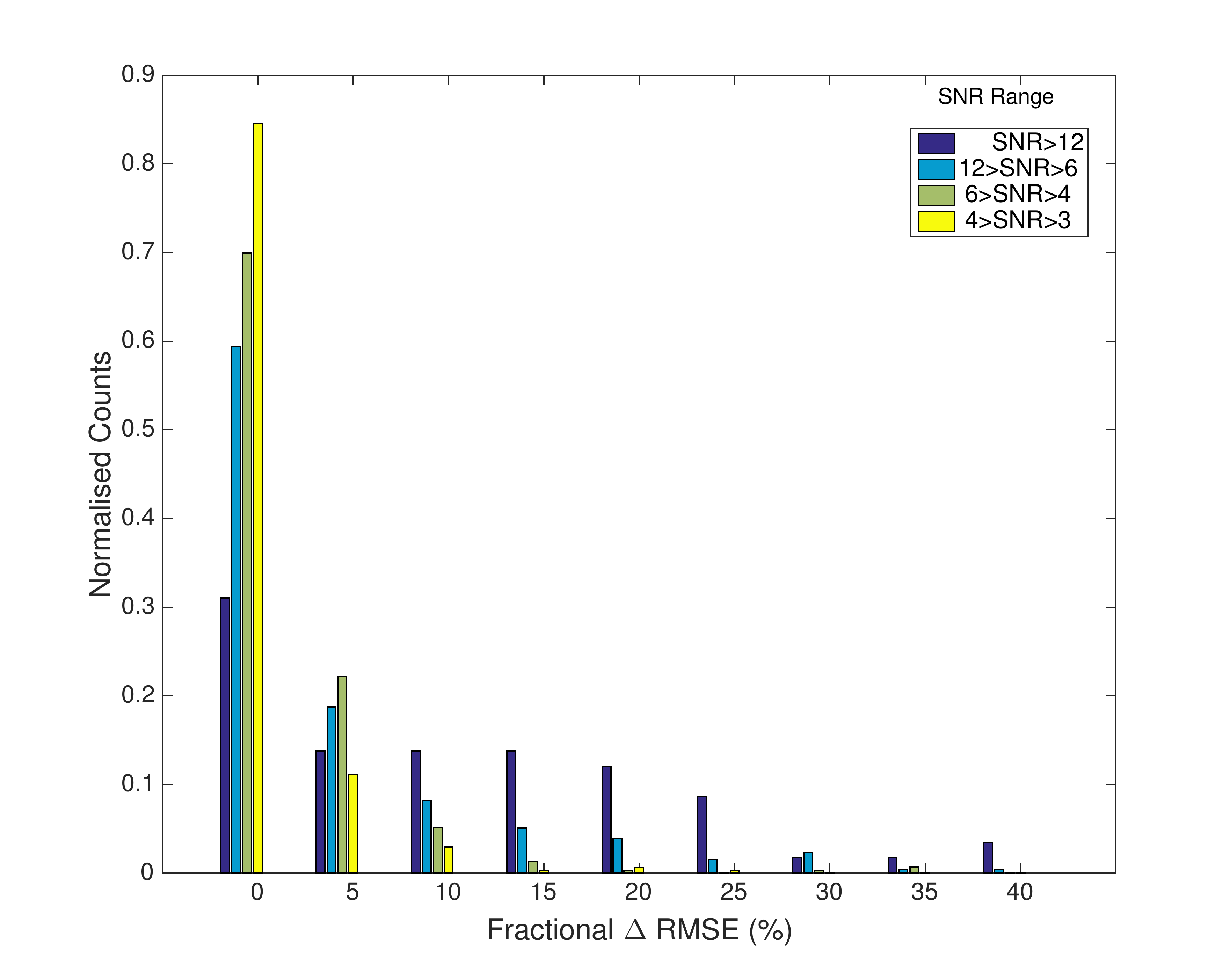}
    \caption{
    The fraction of LEAP ionospheric screens showing significant curvature increases with the SNR per station (derived from the RMSE of the 2D fit residuals). The metric for significant curvature is that the fractional difference of the RMSE values from the linear and second order 2D fits
    is greater than the expectation values from the low signal to noise cases.
    Here we show the normalised counts of directions with ionospheric screen curvatures, for SNR ranges indicated with color codes, at 88 MHz. We conclude that, for SNR$>$6 (or RMSE$<10^o$) most screens are poorly described by a planar surface, even with the small size of MWA-2 array. 
    Therefore, the incidence of curvature is expected to grow larger, in more directions, with increased sensitivity, such as MWA-3, and much more for SKA-Low, which includes much longer baselines, in addition to a giant leap in sensitivity.
    }
    \label{fig:Hist}
\end{figure}

Figure \ref{fig:Hist} shows a histogram of the fractional $\Delta$RMSE values corresponding to the screens shown in Figure \ref{fig:Corrections}, binned by SNR. This shows 
that the ionospheric curvature signature, if present, appears more significant towards the directions of strong LEAP calibrators, where the thermal noise is reduced.






Some limited insights in the temporal behaviour also come from the regression analysis.
We searched the solutions from all the stronger sources for indications of significant variation on shorter time spans than the full scan length of 2min.
Three of the more than 300 source directions with a SNR$>$6 per station, for all scans, at each frequency, showed significant temporal variability. 
Their location is marked on Figure \ref{fig:WSC_corrections}.
Figure \ref{fig:example_phase_temporal} shows a similar plot to Figure \ref{fig:example_phase} for the example with the best SNR, with solutions every 30s. Clear changes in their spatial and temporal structure are easily seen, even by visual inspection. 

{\it Structure Functions:}
Structure function analysis is the traditional method to infer the nature of a propagation media.  
We have formed the spatial structure  functions  of the  ionospheric screens corresponding to the strongest sources in our data across the limited MWA baseline range. For the ``full'' SF we  use the station-based phase values measured with LEAP in the direction of the source, and the values after subtracting the linear plane from the linear regression analysis, for the ``detrended'' SF, respectively. 

\begin{figure}
    \centering
    \includegraphics[width=0.8\textwidth]{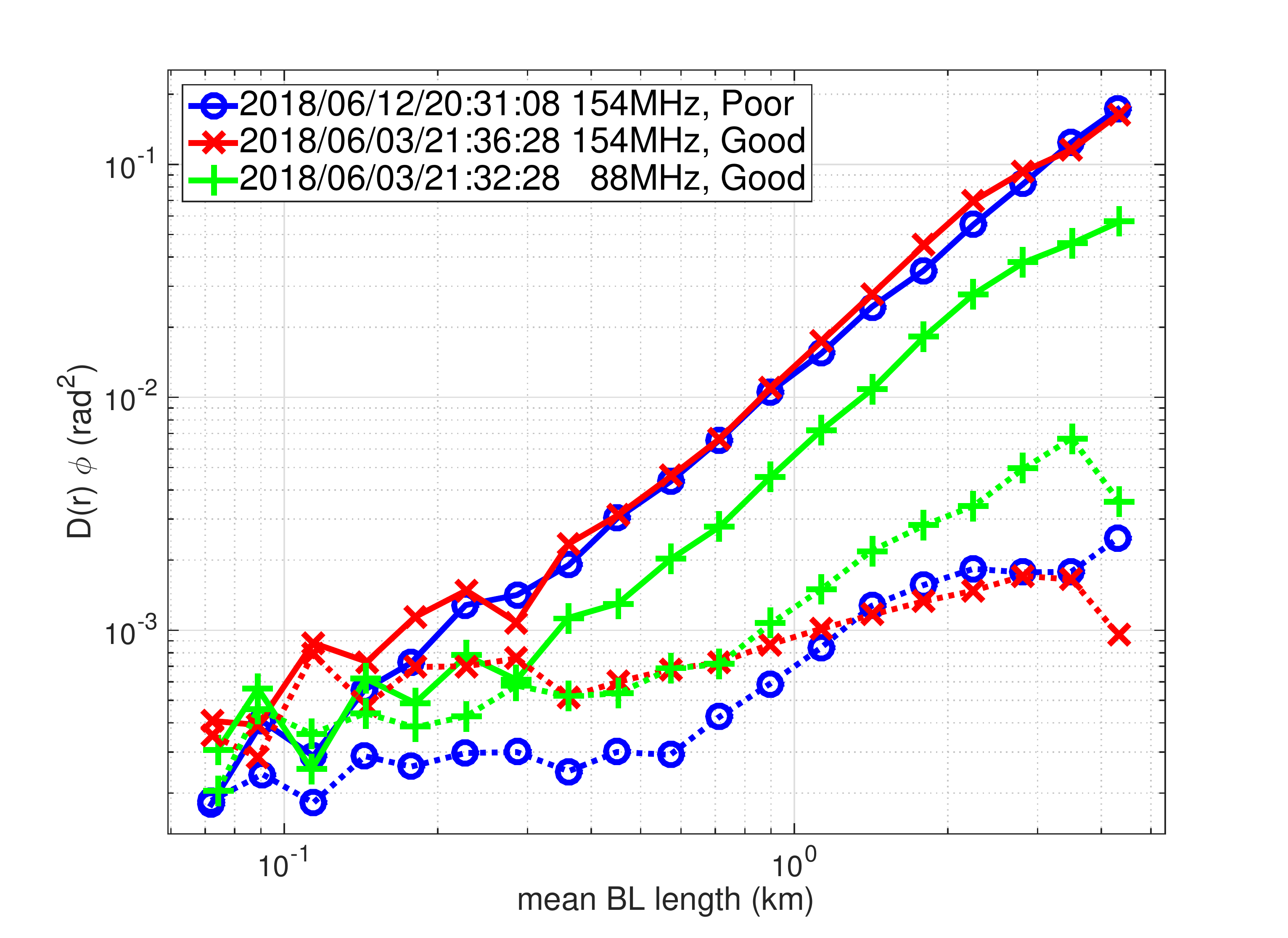}
\caption{Full (solid lines) and detrended (dotted lines) spatial structure function D(r) of the LEAP ionospheric screens measured in the direction of the 
strong source 3C444 (J221425-170140), from MWA-2 observations carried out on June 2018 under good (red crosses) and poor (blue circles) weather conditions at 154MHz, and under good weather conditions (green pluses) at 88MHz. 
The full structure functions follow a near-constant gradient across the baseline range, here shown in units of radians$^2$ for comparison to Mevius \etal{} 2016. 
The detrended structure functions have been calculated after removal of the 2D linear fit from the ionospheric screens.
In this case, the transition from noise dominated domain to baseline length dependent domain occurs at about 0.6km, which would be our recommended scale for the introduction of higher order DD-calibrations. To scale the y-axis to units of mTECU$^2$, for comparison to Figure \ref{fig:SF}, multiply the 88/154 MHz data by 110/338, respectively.}

%
    \label{fig:3C444}
\end{figure}

The clearest example, for the direction towards the strongest source 3C444 with a point source flux of 44\,Jy at 154MHz, is shown in 
Figure \ref{fig:3C444}, for the full (solid lines) and detrended (dotted lines) SF analysis, for observations on two days with ``good'' (red for 154MHz and green for 88MHz) and ``poor'' (blue for 154MHz) weather conditions. 
In this case the SFs increase with baseline length with gradients of 
$1.8^{+0.2}_{-0.2},1.7^{+0.2}_{-0.2}, 1.82^{+0.04}_{-0.04}$, respectively. The noise floor is not reached until very short baselines, below a few hundred meters.
The detrended SF analysis shows a rise of the residual signal over the noise at baselines greater than 0.6km, which is more clear for the ``poor'' weather observations and lower frequencies.

%

\begin{figure}\hspace{-1cm}
    \centering
    \includegraphics[width=0.95\textwidth]{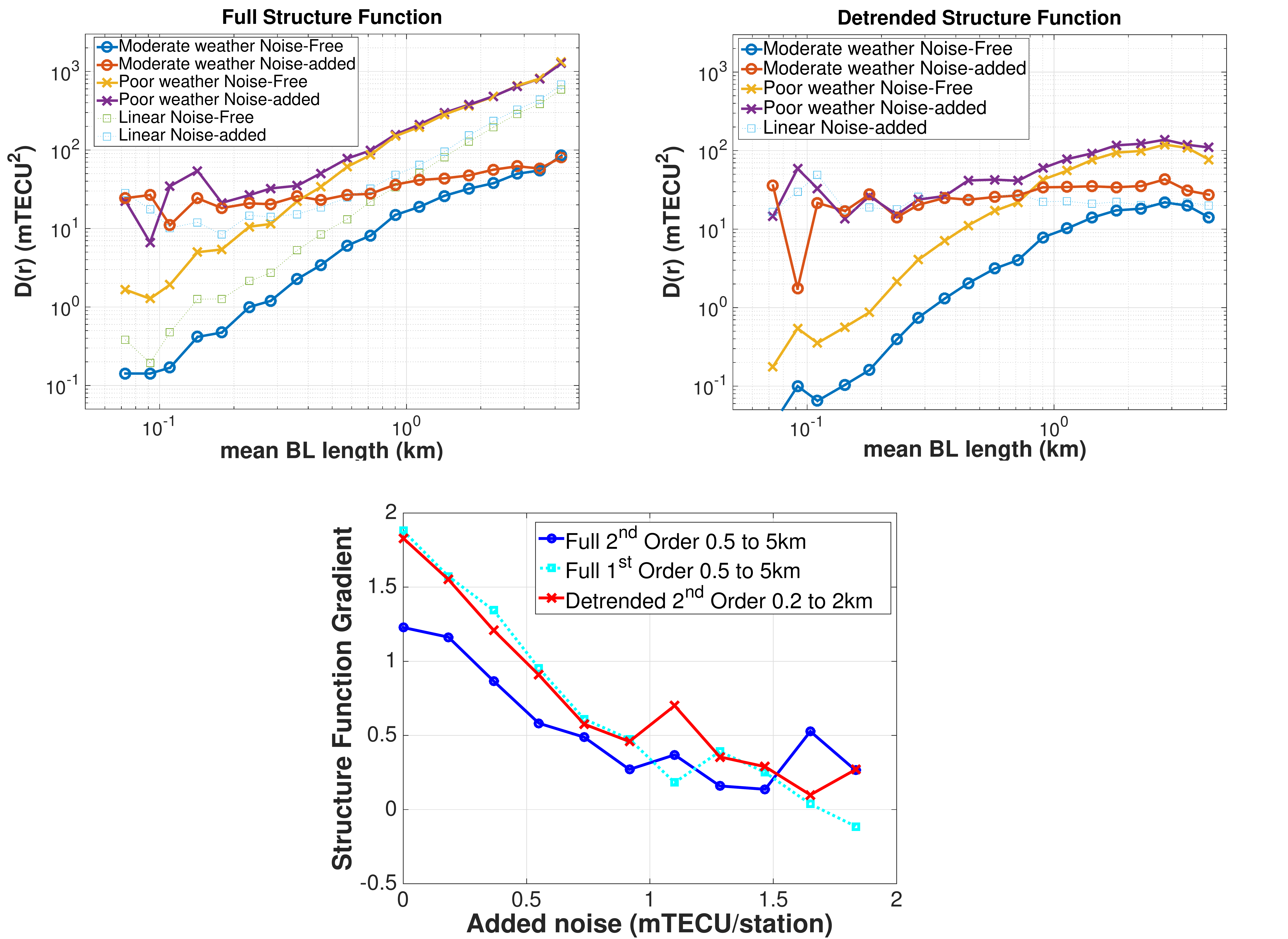}
\caption{Spatial structure functions calculated for simulated toy ionospheric models at 88MHz, based on the ionospheric screens measured with MWA-2. 
{\it Upper Left} shows the traditional full structure function (SF), for a 2D-polynomial model with parameters for curvature and slope typical for our moderate-weather observations (circles), noise-free and with added-noise equivalent to a per station SNR of 20. Plotted with crosses are the SF for poor-weather with parameters approximately four times greater, with and without noise. Finally, with dotted lines and squares are the SF calculated for a planar model, with and without noise.
For a given weather condition, the effect of added noise limit the SF measurements at short baselines and shifts the SF upwards, compared to noise-free case;
however, the values and gradients of the SFs are preserved at longer baselines. 
%
{\it Upper Right} shows the detrended SF, calculated after a 2D linear model has been subtracted from the data describe above, showing the signature of the residual non-linear ionospheric structures. 
We consider any excess signal above the noise-floor region at the shortest baselines, as the indicator for the presence of non-linear screen structure.
The turn-over (or potentially turn up with even higher order terms) towards the longest baselines indicate the end of the inertial region.
{\it Lower} shows the gradients measured from the SF plots verses added noise, for the full SF ({\it blue circles and cyan squares} fitted over the range of 0.5 to 5km and for the detrended SF {\it red crosses} fitted over 0.2 to 2km baseline lengths. The blue line is for a second order screen model and the cyan line is for a planar model. For the latter the noise-free gradient should be exactly 2. 
This shows how the gradient is a sensitive function of the noise-level, particularly with the narrow range of MWA-2 baselines. 
    \label{fig:toy}}
\end{figure}

Figure \ref{fig:toy} shows the behaviour of the spatial structure functions ({\it left} full; {\it right} detrended) for simulated 2D ionospheric screens (i.e. ``toy models'') 
defined by ionospheric slope and curvature values comparable with the measurements from our regression analysis in the 2017 datasets 
(i.e. a slope of 5.1mTECU/km and curvature of 6.8mTECU/km$^2$), to 
four times greater, for moderate (blue dots) and poor (yellow cross) weather,  respectively; for comparison, the case for a linear screen (i.e. no curvature, and a slope double of the minimum) is shown with green squares.
In addition, we tested the effect of measurement noise in the SF by adding a random noise signal with an RMSE of up to 10\deg{} per 
station (i.e. $\sim$2mTECU at 88MHz) to these toy models 
(red circles and purple crosses for moderate and poor weather, respectively). 
Figure \ref{fig:toy} ({\it lower}) shows the behaviour of the gradient of the structure functions, across the limited range of
MWA baseline lengths, as a function of the added noise signals to the toy models.

\begin{figure}
    \centering
    \includegraphics[width=0.95\textwidth]{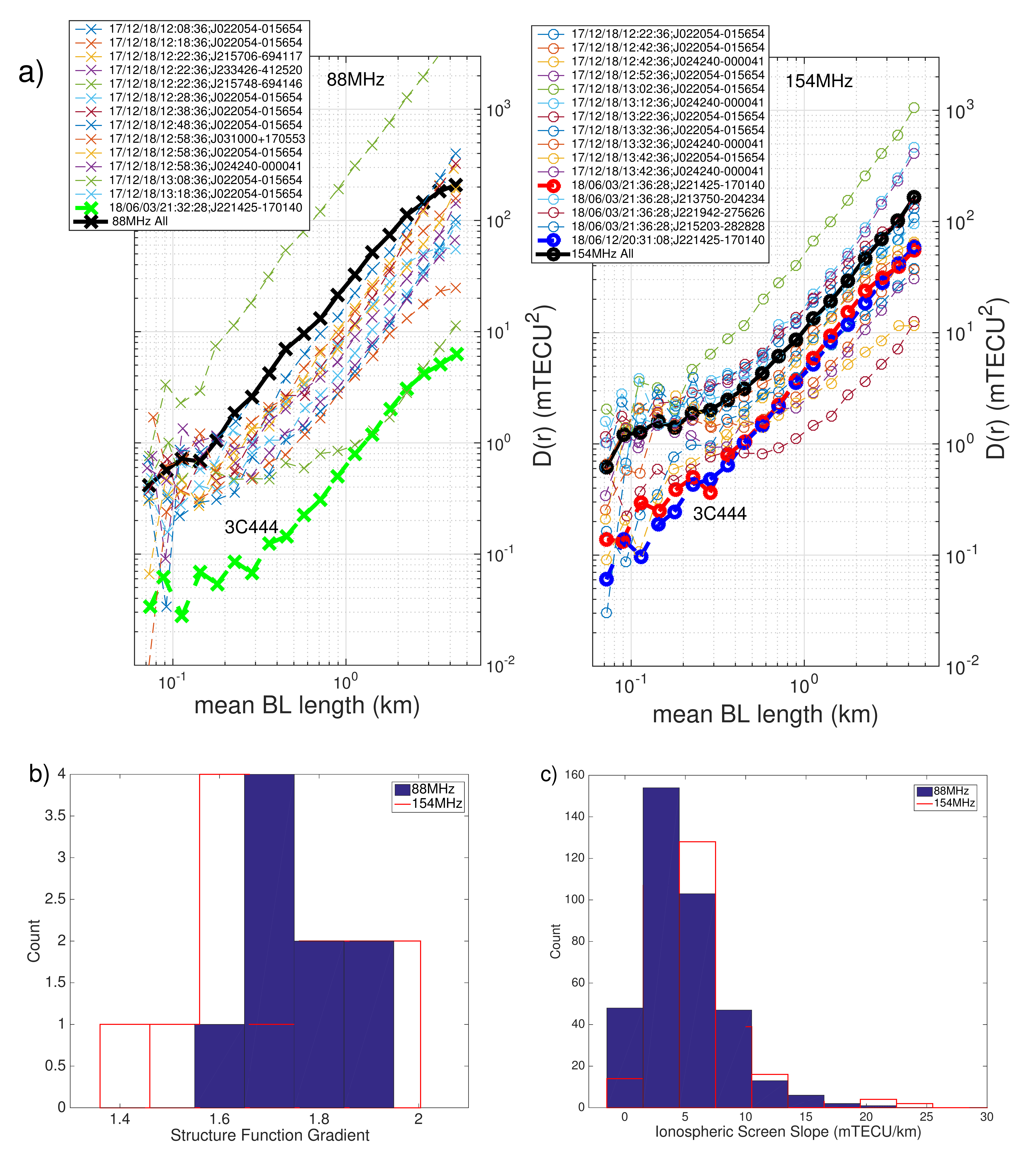}
    \caption{
    {\it Upper plots} show the empirical spatial structure functions calculated from the MWA-2 LEAP ionospheric screens in the directions of sources with SNR $>$20 (per station) versus projected baseline bins at 88 ({\it left}) and 154 MHz ({\it right}). The intermittent lines are for measurements of an individual screen, colour coded to identify the time of the 2-min scan and source direction. The bold thick black line comprises the measurements from the ensemble of all screens, that is multiple scans and directions, to calculate the structure function. 
    The bold coloured lines, labelled with 3C444, are repeats of the data from Figure \ref{fig:3C444}.
    The structure functions are shown in units of mTECU$^2$ to allow for direct comparison between 88 and 154 MHz. 
    For the case of lowest measurement noise individual screens (towards 3C444, with SNR$>$50, shown with the same bold colours as in Fig. \ref{fig:3C444}) the SF extends in a quasi linear fashion to baselines of $\sim$100m. 
    b) A histogram of the SF gradients in a) measured between 0.5 and 5\,km, plotted in blue solid for 88MHz and red outlines for 154MHz. The median gradients is 1.72$\pm$0.15, which is consistent within errors to that of Mevius \etal{} (2016\cite{mevius_16}) and the expectations for a 2D Kolmogorov spectrum (1.89$\pm$0.1 and 1.67,  respectively).
    c) A histogram of the empirical MWA-2 LEAP ionospheric screen slopes $C$ for all the LEAP calibrators from the regression analysis, in mTECU/km, at 
    88MHz and 154MHz. The median slope is 5$\pm$3mTECU/km at both frequencies. 
    %
    \label{fig:SF}}
\end{figure}
\begin{figure}
    \centering
    \includegraphics[width=0.95\textwidth]{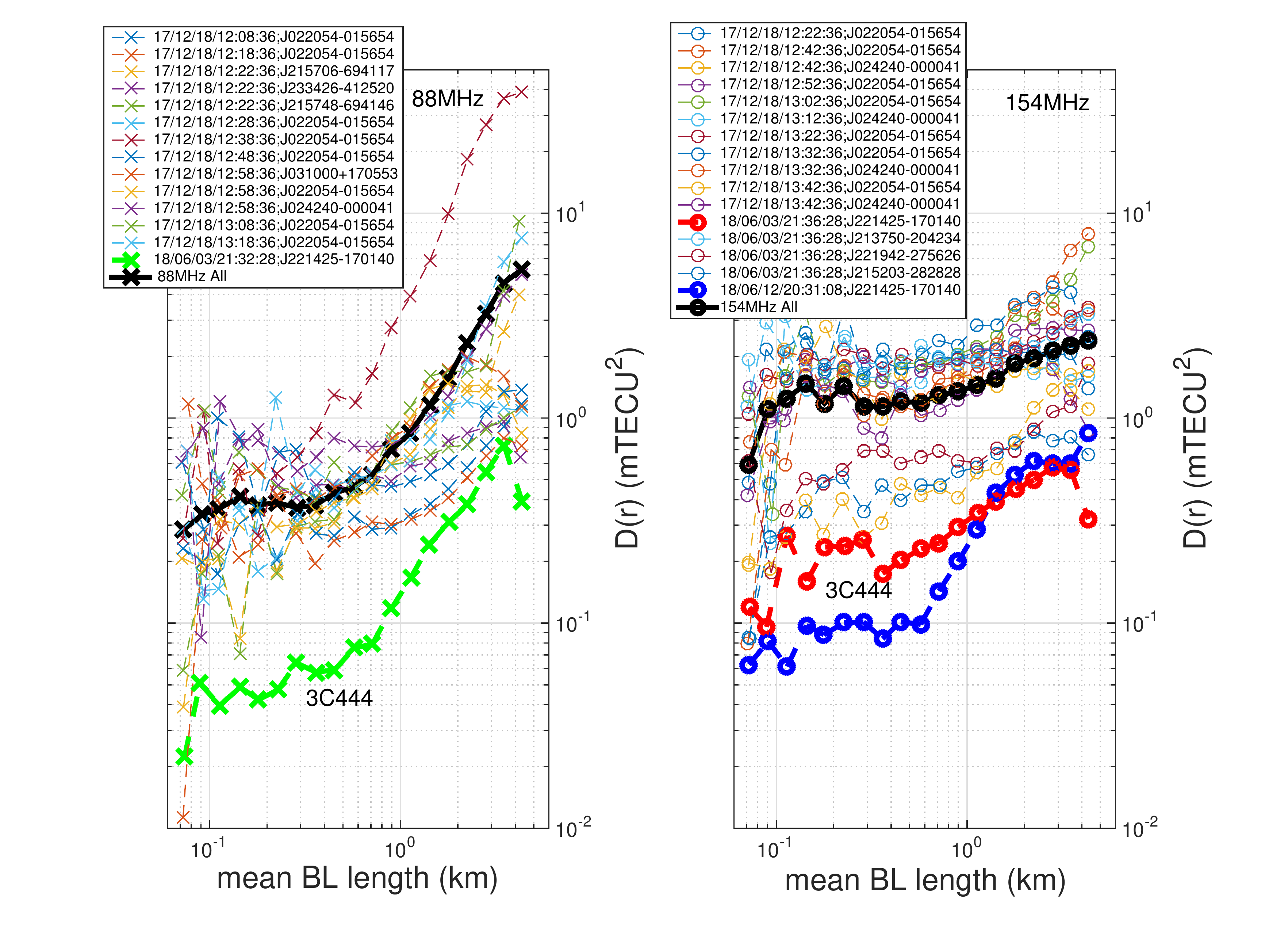}
    \caption{
    \label{fig:SF-detrend} 
    Empirical detrended spatial structure functions calculated from the MWA-2 LEAP ionospheric screens in the directions of sources with SNR $>$20 (per station) after subtracting a planar fit versus projected baseline bins at 88 ({\it left}) and 154 MHz ({\it right}). 
    Any excess signal above the noise unambiguously exposes the presence of the `non-linear' components in the ionospheric screens.
    The DSF for individual screens in the direction of 3C444 are shown in bold (green, blue, red) matching the colours in Fig.  \ref{fig:3C444}, and for the ensemble over time and directions, shown in bold and black.
    The bold coloured lines, labelled with 3C444, are repeats of the data from Figure \ref{fig:3C444}.
    The ensemble DSF at 154MHz does not show significant non-linear terms greater than the noise, although individual datasets do show such behaviour. On the other hand, both the ensemble and individual detrended structure functions at 88MHz indicate the presence of significant non-linear structure at small scales, i.e. baseline lengths greater than 0.6km.}
    
\end{figure}

Figure \ref{fig:SF} shows a compilation of the full structure functions of the ionospheric screens towards strong (SNR$>$20) sources, for each individual 2-min measurement set and source direction
({\it dashed lines}).
Also using multiple scans and directions to form the combined SF (black solid line, in bold), at 88 MHz ({\it upper left}) and 154 MHz ({\it upper right}). 
%
A histogram of the structure function gradients over the range of 0.5 to 5 km baselines is shown ({\it lower left}),
along with the histogram of the ionospheric slopes (measured on the input ionospheric screens with the linear regression analysis ({\it lower right}).
%
Figure \ref{fig:SF-detrend} shows the detrended structure functions for the same datasets as Figure \ref{fig:SF}, 
at 88 MHz ({\it left}) and 154 MHz ({\it right}); 
here, after the contribution from a nominal planar surface is subtracted out, the detrended SF is dominated by the presence of non-linear structures in the ionospheric screens.

\section{Discussions and Conclusions}

%

\subsection{Regression Analysis}

We have used the LEAP measurements of the ionospheric phase surfaces over the MWA-2 array to make a 
detailed analysis of the spatial structure of the ionospheric distortions imposed on the incoming wavefronts towards $\sim$200 simultaneous directions across the FoV, for 20 datasets spanning a range of different weathers, at 154 and 88 MHz.
LEAP, being an station-based phase calibration, is sensitive to higher order ionospheric phase structural changes at small scales, less than 
the size of the array.
Alternative calibration strategies that use image-domain apparent source position shifts to measure the ionospheric phases only measure a simple gradient\cite{loi_15,jordon_17,helmboldt_20}. 
However, LEAP analysis requires observations of stronger sources, as we now are determining antenna-based corrections\cite{rioja_18}, rather than corrections from the array-averaged data in the image.
Nevertheless, ignoring higher order effects results in 
severe artefacts in the images and introduces an intractable bias in studies such as EoR\cite{vedantham_15,trott_18}. 
Here we discuss the results of two approaches to measure spatial small scale phase deviations 
which are undetectable to analysis based on the image-domain position shifts. 


We are confident that the LEAP calibration method is providing measurements of only the ionospheric surface. Standard practise in MWA analysis is to assume that the beam response is the same for all stations. If there were deviations from this assumption random distortions would appear in the phase surfaces. Such behaviour would be expected to occur mainly on the short baselines, where the stations can interfere with each other. Given that we don’t see such behaviour, i.e. the surfaces are smooth, we can take that the assumption is valid, particularly for this extended baseline configuration.

For the regression analysis  
we fitted 1$^{\rm st}$ and 2$^{\rm nd}$ order polynomial to the surfaces to estimate the ionospheric screen parameters: slope, direction and curvature. 
These display large scale coherent behaviour in the slopes (Figure \ref{fig:Corrections}), over the FoV, as previously reported \cite{rioja_18,jordon_17,helmboldt_20}.
The RMSE values for the 2$^{\rm nd}$ order fit correlates well with the inverse of the LEAP calibrator flux densities and the expected thermal SNR; that is, the stronger sources have  smaller fitted parameter errors, 
as expected from high signal to noise ratio measurements.
The coherent behaviour in the curvatures, derived in the 2$^{\rm nd}$ order fit, over the sky is less clear because of the sparsity of strong calibrator signals, but does indicate spatial coherence (Figure \ref{fig:Curved_corrections}) at the lower frequency.

The fractional RMSE change between the 1$^{\rm st}$ and 2$^{\rm nd}$ order fits provides a useful metric for the significant detection of curvature in the presence of measurement noise. 
%
When considering all observations, the RMSE values from the first and second order fits are not 
significantly different for the majority of the cases. This is to be expected, as the small collecting area 
of the MWA stations and the low sensitivity result in a thermal noise contribution larger than the ionospheric phase signature 
for the majority of LEAP calibrator sources. 
Figure \ref{fig:Hist} shows that, when introducing  LEAP calibrator flux density cutoffs, the impact of the curvature is increasingly visible 
for higher SNRs  
with 20\% and 40\% of cases showing significant ($>\sim 15$\%) fractional RMSE improvements, 
for 6$<$SNR$<$12 and SNR$>$12, respectively.

The presence of curvature in the ionospheric screens over the array, if uncorrected, result in increased residual sidelobes after deconvolution, particularly for stronger sources.
%
This underlines the importance of station-based calibration to reduce the residual sidelobes in the cleaned images.
This will be fully discussed in Dodson \etal{} (2021 in prep.), where we show the 
improvement in the recovered source peak fluxes in the images.
We have limited information for the discussion of high temporal variability of the phase screens, other than noting that three out of the about three hundred strong (SNR$>$6) sources in all the datasets showed significant variation at short timescales. One example is plotted in Figure \ref{fig:example_phase_temporal}. Because of the few lines of sight towards strong sources it is impossible to track the nature of this behaviour, which will affect the more sensitive SKA observations.

Significant small scale deviations from linearity in the spatial ionospheric phase distributions are detected from the regression analysis, 
even with the limited sensitivity and size of MWA Phase-2. 
This indicates that MWA Phase-2 observations reside in the Lonsdale regime 4. 
Under these conditions, the performance of image-based apparent source position shift ionospheric calibration degrades, 
thus there will be benefits from using higher order calibration with MWA Phase-2. 
These effects are expected to become more significant in observations with MWA Phase-3, and even more with SKA, as discussed below.

\subsection{Structure Function Analysis}



Because of the excellent ground coverage of the MWA stations, our observations are highly applicable for SF analysis, to characterise
the fine scale structure of the ionospheric wavefront distortions above the array towards multiple viewing directions, and probe the underlying physical nature of the distorting media. 

%
Because of the denser station coverage we are able to reconstruct the fine structure much better than the similar study on LOFAR\cite{mevius_16}, whilst the latter explores a much large range of baseline lengths. The observational differences lead to slightly different methods; for example the tracking LOFAR observations allow for the subtraction of the temporal mean phase, where as for the snapshot MWA observations we have subtracted a DI calibration. 
Because of this, 
in cases where one very strong source dominates the DI calibration, the residual ionospheric screen slope in the direction of that source (e.g. 3C444) can be significantly lower than the median slope over all directions. 
An example of this can be seen in Figure \ref{fig:3C444} where the intercept at 1\,km (that corresponds to the slope of the ionospheric screen) of the SF at 88\,MHz is less than the intercept of the 154\,MHz SF, where prima-facie one would expect the reverse. This occurs because of the greater dominance of this source in the DI calibration at the lower frequency. However, the log-log SF gradients are preserved. 
The benefits from high signal-to-noise ratio measurements are the potential to probe smaller scales, increased precision of ionospheric calibration, and tighter constraint of the upper limit of the inner scale of turbulence, in the absence of other systematic errors.

MWA is also free of potential clock errors, which were one of the contributions to the systematic noise floor limits in the LOFAR analysis;
for 3C444 we reach a noise floor of 0.1, 0.3 and 0.05 mTECU
for our three datasets in poor and good weather at 154MHz and good weather at 88MHz, as measured in the detrended data over the baseline range of 50 to 100m.
Nevertheless the two approaches reach similar results as to the SF gradients being most compatible with Kolmogorov turbulence. Mevius \etal{} use the `refractive scale' $R_{\rm diff}$ as a physically relevant measurement (and a proxy of ionospheric weather quality), whereas we use the screen slope (mTECU/km) which is independent of the SF gradient. These two quantities are trivially convertible and comparable between the two studies. 
Our median slopes of 5$\pm$3mTECU/km is equivalent to $R_{\rm diff}$ being 2 to 12km, and  $R_{\rm diff}$ values of more than 5\,km are considered suitable for EoR observations\cite{mevius_16}; this of course assumes that the impact of severe weather manifested as a high ionospheric slope is also turbulent, and ignores the potential for a high quality DD-calibration to turn a nominally `bad weather day' into a good weather day.
In our comparison of the range of $R_{\rm diff}$ we note the differences in the methods; LOFAR is a pointed tracking observation with a constant ionospheric wedge over the array subtracted, whereas this analysis is for a wide-FoV experiment with many lines of sight where a global DI model has been subtracted. These differences will affect the deduced $R_{\rm diff}$, making exact comparison difficult. 
%

We have used toy surface models to improve our interpretations of the 
SF method. 
These show that in the presence of noise, interpreting a gradient of $\sim$5/3 as proof of Kolmogorov turbulence is risky, particularly with the short baseline range of the MWA, as noise flattens the gradient in the full structure function. 
We have investigated the impact of joint fitting of the noise floor plus the noise-free structure function (i.e. $C^2r^\beta + \sigma^2$), but found that for the MWA the terms $\beta$ and $\sigma$ are highly correlated. 
We found our best results by limiting the measurement of the gradient to the range of baselines least affected by the noise (that is 0.5 to 5km).
Finally, the signature of the toy model's curvature is hard to see in the SF, as it is dominated by the linear component.
However, the detrended structure function does allow for the robust detection of non-linear distortions, the spatial scales for the onset of these effects and for the end of the inertial-like regime. 

%

For the strong source 3C444 (Figure \ref{fig:3C444}) we can track the structure functions down to the shortest baselines without being overcome by the thermal noise; the gradient in this case is 1.8 at 154\,MHz in both good and poor weather days, and 1.7 at 88\,MHz, between 0.5 and 5km. 
Thus at both frequencies the best fit interpretation would agree with being Kolmogorov turbulence, but we don't consider this proven, given the difficulty of separating the gradient from the noise. 
%
%
Using DSF (i.e. after subtracting a linear surface) the detectable divergence from the linear fit occurs at less than 1km, which we can confidently interpret as the first indications of non-linear behaviour above the noise. 
At the longest baselines there is a change in the detrended structure function gradient, which is characteristic of curvature. 
In summary, we would recommend that non-linear corrections are applied for baselines greater than 1km.

Similar conclusions come from the study of the ensemble of directions and times to strongest sources in all the datasets; 
the gradients are consistent with Kolmogorov turbulence, being 1.6$\pm$0.2 and 1.8$\pm$0.1 for the 88MHz and 154MHz datasets (Figure \ref{fig:SF}). But we stress that these measured gradients could be underestimated because of the impact of the thermal noise.
The detrended SF (Figure \ref{fig:SF-detrend}) allow us to better detect deviations from a planar surface.
These show that in general at 88MHz, even in moderate-weather, higher order ionospheric corrections should be used for baselines greater than 0.6km and must be used for baselines greater than 2km.
The combined SF at 154MHz are more consistent with planar solutions, except for strong sources (i.e. as shown in Figure \ref{fig:3C444}) and/or poor weather, when higher order corrections are required over the same baseline ranges.
This implies that MWA-2 falls in Lonsdale's regime 4.

\subsection {Implications for SKA}

The SKA-Low stations will be approximately ten times larger, thus have close to a hundred times larger collecting area (per station). Furthermore the instantaneous bandwidth will be ten times larger. 
Therefore the sensitivity will be much higher than for MWA, and the importance of precise ionospheric calibration (and other systematic errors), will be even more significant. 
The expected continuum baseline sensitivity is 3mJy in 1 minute\cite{ska_perf}, from which we predict more than 100 sources with SNR$>$6 in the FoV, using the TREC models\cite{TREC}.

We have found that for MWA at 88MHz, even in moderate-weather, the influence of higher order terms are detectable for baselines greater than 0.6km and are significant for baselines greater than 2km.
On the other hand, for most sources at 154MHz in moderate-weather planar solutions should be sufficient, but for strong sources (i.e. as shown in Figure \ref{fig:3C444}) and/or poor weather (or both) higher order corrections are required. 
These conclusions will be most applicable for the SKA, where the increase in sensitivity mean that many sources will be at similar or higher SNR that the few limited examples with the MWA. 
Thus we conclude that station based directional dependent calibration solutions are required for both the final imaging, 
and also for the tied-array beam forming, even for $\sim$kilometre-long baselines. 
%
The VLBI beamforming, for example, is assuming that all stations with in a radius of 20\,km (or as a fall-back position 4\,km) will be included in the formed tied-array beams\cite{skavlbi_wp10}.
In this case real-time station-based direction-dependent calibration will be required; this can be provided by LEAP, as we are currently working on demonstrating.

In summary the opportunities opened up by the amazing potential of SKA-Low to revolutionise low-frequency astronomy require that the calibration of the data is equally accurate. Thus station-based directional-dependent corrections will be needed for all baselines and operations.

\vspace{2ex}\noindent\textbf{Mar\'ia J. Rioja Capellan} is an senior research fellow with joint appointments at the University of Western Australia, Space And Astronomy, CSIRO (AU) and the Observatorio Astron\'omico Nacional (IGN-OAN, ES) with 
more than 60 journal papers. Her research focus is mainly in astrophysical applications of high-precision astrometry, and thus the correction of distortions on long baselines to radio interferometric data. She is focusing on the development of the next-generation of astronomical methods to allow for the delivery on the power of the next-generation of radio telescopes. 

\vspace{2ex}\noindent\textbf{Richard Dodson} is an senior research fellow at the University of Western Australia (AU).
His research focus is on the observational issues the SKA must address to deliver on the potential of this ground-breaking instrument,
thus the systematic errors that will be the true limitations. These include those that come from: the ionosphere for SKA-Low, the atmosphere for SKA-VLBI, and imperfections in imaging methods for deep spectral line imaging. 


\listoffigures
\listoftables


\begin{thebibliography}{10}

\bibitem{sovers_98}
O.~J. {Sovers}, J.~L. {Fanselow}, and C.~S. {Jacobs}, ``{Astrometry and geodesy
  with radio interferometry: experiments, models, results},'' {\em Reviews of
  Modern Physics} {\bf 70}, 1393--1454  (1998).

\bibitem{lonsdale_05}
C.~J. {Lonsdale}, ``{Configuration Considerations for Low Frequency Arrays},''
  in {\em From Clark Lake to the Long Wavelength Array: Bill Erickson's Radio
  Science},  N.~{Kassim}, M.~{Perez}, W.~{Junor}, {\em et~al.}, Eds., {\em
  Astronomical Society of the Pacific Conference Series} {\bf 345}, 399
  (2005).

\bibitem{ddf_tasse}
C.~{Tasse}, B.~{Hugo}, M.~{Mirmont}, {\em et~al.}, ``{Faceting for
  direction-dependent spectral deconvolution},'' {\em A\&A} {\bf 611}, A87
  (2018).

\bibitem{offringa_14}
A.~R. {Offringa}, B.~{McKinley}, N.~{Hurley-Walker}, {\em et~al.}, ``{WSCLEAN:
  an implementation of a fast, generic wide-field imager for radio
  astronomy},'' {\em \mnras} {\bf 444}, 606--619  (2014).

\bibitem{tol_18}
S.~{van der Tol}, B.~{Veenboer}, and A.~R. {Offringa}, ``{Image Domain
  Gridding: a fast method for convolutional resampling of visibilities},'' {\em
  \aap} {\bf 616}, A27  (2018).

\bibitem{intema_09}
H.~T. {Intema}, S.~{van der Tol}, W.~D. {Cotton}, {\em et~al.}, ``{Ionospheric
  calibration of low frequency radio interferometric observations using the
  peeling scheme. I. Method description and first results},'' {\em A\&A} {\bf
  501}, 1185--1205  (2009).

\bibitem{mitch_cal}
D.~A. {Mitchell}, L.~J. {Greenhill}, R.~B. {Wayth}, {\em et~al.}, ``{Real-Time
  Calibration of the Murchison Widefield Array},'' {\em IEEE Journal of
  Selected Topics in Signal Processing} {\bf 2}, 707--717  (2008).

\bibitem{sagecal}
S.~{Yatawatta}, F.~{Diblen}, H.~{Spreeuw}, {\em et~al.}, ``{Data multiplexing
  in radio interferometric calibration},'' {\em \mnras} {\bf 475}, 708--715
  (2018).

\bibitem{skavlbi_wp10}
C.~Garcia-Miro, ``{WP10: VLBI with the SKA},'' tech. rep., JIVE  (2019).

\bibitem{nhw_gleam}
N.~{Hurley-Walker}, J.~R. {Callingham}, P.~J. {Hancock}, {\em et~al.},
  ``{GaLactic and Extragalactic All-sky Murchison Widefield Array (GLEAM)
  survey - I. A low-frequency extragalactic catalogue},'' {\em \mnras} {\bf
  464}, 1146--1167  (2017).

\bibitem{mevius_16}
M.~{Mevius}, S.~{van der Tol}, V.~N. {Pandey}, {\em et~al.}, ``{Probing
  ionospheric structures using the LOFAR radio telescope},'' {\em Radio
  Science} {\bf 51}, 927--941  (2016).

\bibitem{loi_15}
S.~T. {Loi}, C.~M. {Trott}, T.~{Murphy}, {\em et~al.}, ``{Power spectrum
  analysis of ionospheric fluctuations with the Murchison Widefield Array},''
  {\em Radio Science} {\bf 50}, 574--597  (2015).

\bibitem{jordon_17}
C.~H. {Jordan}, S.~{Murray}, C.~M. {Trott}, {\em et~al.}, ``{Characterization
  of the ionosphere above the Murchison Radio Observatory using the Murchison
  Widefield Array},'' {\em \mnras} {\bf 471}, 3974--3987  (2017).

\bibitem{helmboldt_20}
J.~F. {Helmboldt} and N.~{Hurley-Walker}, ``{Ionospheric Irregularities
  Observed During the GLEAM Survey},'' {\em Radio Science} {\bf 55}, e07106
  (2020).

\bibitem{rioja_18}
M.~J. {Rioja}, R.~{Dodson}, and T.~M.~O. {Franzen}, ``{LEAP: an innovative
  direction-dependent ionospheric calibration scheme for low-frequency
  arrays},'' {\em \mnras} {\bf 478}, 2337--2349  (2018).

\bibitem{gleam-x}
N.~{Hurley-Walker}, N.~{Seymour}, L.~{Staveley-Smith}, {\em et~al.},
  ``{GaLactic and Extragalactic All-Sky MWA-eXtended (GLEAM-X) survey: Pilot
  observations}.'' MWA Proposal id.2017A-11  (2017).

\bibitem{mwa2}
R.~B. {Wayth}, S.~J. {Tingay}, C.~M. {Trott}, {\em et~al.}, ``{The Phase II
  Murchison Widefield Array: Design overview},'' {\em PASA} {\bf 35}, e033
  (2018).

\bibitem{nhw_pc}
N.~{Hurley-Walker}, ``{Characterisation of the Ionosphere via Temporal
  Variation of Antenna Gains}.'' MWA memo  (2019).

\bibitem{vantol_phd}
S.~Van~der Tol, {\em {Bayesian estimation for ionospheric calibration in radio
  astronomy}}.
\newblock PhD thesis, University of Delft  (2009).

\bibitem{nature_sf_swarm}
P.~{De Michelis}, G.~{Consolini}, A.~{Pignalberi}, {\em et~al.}, ``{Looking for
  a proxy of the ionospheric turbulence with Swarm data},'' {\em Scientific
  Reports} {\bf 11}, 6183  (2021).

\bibitem{sf_toys_neuroscience}
R.~A. Antonia and C.~W.~V. Atta, ``Structure functions of temperature
  fluctuations in turbulent shear flows,'' {\em Journal of Fluid Mechanics}
  {\bf 84}(3), 561–580  (1978).

\bibitem{vedantham_15}
H.~K. {Vedantham} and L.~V.~E. {Koopmans}, ``{Scintillation noise in widefield
  radio interferometry},'' {\em \mnras} {\bf 453}, 925--938  (2015).

\bibitem{trott_18}
C.~M. {Trott}, C.~H. {Jordan}, S.~G. {Murray}, {\em et~al.}, ``{Assessment of
  Ionospheric Activity Tolerances for Epoch of Reionization Science with the
  Murchison Widefield Array},'' {\em ApJ} {\bf 867}, 15  (2018).

\bibitem{ska_perf}
R.~Braun, A.~Bonaldi, T.~Bourke, {\em et~al.}, ``Anticipated performance of the
  square kilometre array -- phase 1 (ska1),''  (2019).

\bibitem{TREC}
A.~{Bonaldi}, M.~{Bonato}, V.~{Galluzzi}, {\em et~al.}, ``{The Tiered Radio
  Extragalactic Continuum Simulation (T-RECS)},'' {\em \mnras} {\bf 482}, 2--19
   (2019).

\end{thebibliography}
\end{document}